\documentclass[12pt]{article}
\usepackage{amsmath,amsthm,amsfonts,amssymb,amscd}
\usepackage{graphics}
\usepackage[latin1]{inputenc}

\renewcommand{\thefootnote}

\begin{document}

\vglue .3in

\centerline{\large\bf On loop space-self avoiding string representations}

\centerline{\large\bf  for $\pmb{Q.C.D(SU(\infty))}$}

\vglue .4in

\centerline{\large\bf Luiz C.L. Botelho}

\vglue .3in

\centerline{Departamento de Matemática Aplicada}

\smallskip

\centerline{Instituto de Matemática, Universidade Federal Fluminense}

\smallskip

\centerline{Rua Mario Santos Braga, CEP 24220-140}

\smallskip

\centerline{Niterói, Rio de Janeiro, Brasil}

\smallskip

\centerline{e-mail: botelho.luiz@superig.com.br}

\vglue .5in

\noindent{\large\bf Abstract:}\, We present several clarifying comments on the loop space-self avoid string representation for $Q.C.D(SU(\infty))$ proposed by this author along last decades.

\bigskip


\vglue .4in

\noindent{\large\bf 1.1\, Introduction}

We star our introduction by recalling the following higlights from eminent Physicists on the search for a mathematical formalism free from ambiguities in strong and weak strong nuclear forces theory
\medskip

\begin{itemize}
\item[{1-}] ``Therefore conclusions based on the renormalization group arguments concerning Q.F.T summed to all orders are dangerous and must be viewed with due caution. So is it with all conclusion from Q.C.D."
\end{itemize}
\begin{itemize}
\item[{2-}] ``Because of severe divergences, Yang-Mills theories can not be consistently interpreted by conventional perturbation (LSZ) theory."
\item[{3-}] ``There are method and formulae in science, which serve as master-key to many apparently different problems. The resource of such things have to be refilled from time to time. In A.M. Polyakov's opinion (and mine), we have to develop a formalism to sum over random surfaces and Riemann geometries (``Brownian Riemann Surfaces")".
\end{itemize}

These comment are intend to clarify the concept of self-avoiding string representation for Q.C.D in the formalism of Loop Space Theory, mostly  exposed in our previous works on the subject ([1],[2],[3]). A word about this comment writing format. Since loop space $Q.C.D(SU(\infty))$ and self-avoiding string path integrals are a notorously difficult mathematical methods subject to be exposed, we deliberately have transferred to 4 appendixes those more mathematical oriented arguments supporting the text main discussions.

\vglue .4in

\noindent{\large\bf 1.2\, Revisiting the loop space formulation for Q.C.D.}

\vglue .2in

Since its proposal several decades ago, the loop space formulation of Q.C.D has been an alternative to the well-known Feynman-diagramatics-perturbative formulation of well-defined quantum field theories ([2],[5]).

In this section we intend to point out some the mathematical difficulties on applying the loop space formalism for Q.C.D or to any other quantum field theory with spinorial matter field.

Let us write the generaling functional of color singlet bilinear vectorial curents with the quark fermionic degrees integrated out
\begin{equation}
Z[J_\mu] = \frac{1}{Z(0)}\, \bigg\langle \det \not\!\!{D}(A_\mu+J_\mu)\bigg\rangle_{SU{(N_c)}} \tag{1}
\end{equation}
Here $\langle \rangle_{SU(N_c)}$ denotes the formal average of the quantum Yang-Mills fields on $SU(N_c)$.

The next step is write the above functional determinat as a continuous sum over the massless quark trajectories. In a plane wave (euclidean) spinor bases
\begin{equation}
\langle x,\alpha\rangle = e^{ipx}\,U_\alpha^{(1)}(p) \tag{2-a}
\end{equation}
\begin{equation}
\overline{(y,\beta\rangle} = \overline{U}_\beta^{(2)}\,e^{ipy} \tag{2-b}
\end{equation}
one has the Feynman path integral formal expression for the above written quark determinant
\begin{align*}
\ell n(\det\not{\!\!D}(A+J)) 
&= -\left\{\int_0^\infty \frac{dt}{t}\,{\rm Tr}\big(e^{-(\not{\!\!D}(A+J))}\big)\right\}\\
&=-\bigg\{ \int_0^\infty \frac{dt}{t} \int d^Dx_\mu\,d^D p_\mu \bigg\{\int_{X^\mu(0)=X^\mu(t)=x^\mu} D^F[X^\mu[(\sigma)] \times \int_{P^\mu(0) = P^\mu(T) = p^\mu}\\
&\,\exp\left(i\int_0^t P_\mu(\sigma)\dot{X}^\mu(\sigma)\,d\sigma\right)\\
&\,\mathbb{P}_{\rm Dirac}\,\mathbb{P}_{SU(N)} \left[\exp\left(i\int_0^t \gamma^\mu\left((P_\mu(\sigma) + A_\mu(X(\sigma)) + J_\mu(X(\sigma)\right)\right)\right]\bigg\} \tag{3}
\end{align*}

Unfortunatelly the path integral object eq(3) remains not well understood from a mathematical point of view as far as this author knows.

However, one can use a formal path $SU(N)$ valued variable change: 
\begin{equation}
P_\mu(\sigma) + A_\mu(X(\sigma)) + J_\mu(X(\sigma)) = \pi_\mu(\sigma) \tag{4}
\end{equation}

One thus gets the ``less formal" and more mathematical expression palatable from a Theoretical Physics point of view for the fermion determinant
\begin{align*}
&\ell n\big(\det \not{\!\!D} (A_\mu+J_\mu)/\det 
\not{\!\!D}(0)\big)\\
&=-\int_0^\infty \frac{dt}{t} \bigg[\int_{X^\mu(0)=X^\mu(t)} D^F[X(\sigma)\bigg]\\
&\,\,\int_{\pi^\mu(0)=\pi^\mu(t)} D^F[\pi^\mu(\sigma)] {\rm Tr}_{SU(N)}\,{\rm Tr}_{\rm Dirac}\,\mathbb{P}_{\rm Dirac}\,
\mathbb{P}_{SU(N)}\\
&\times\bigg\{\exp \left(+i\int_0^t (\gamma^\mu \pi_\mu)(\sigma)d\sigma\right)\\
&\,\,\times {\rm Tr}_{SU(N)} \left[\mathbb{P}_{SU(N)}\exp\left(-i\int_0^t (A^\mu(X(\sigma))-\pi^\mu(\sigma))dX_\mu(\sigma)\right)\right]\\
&\,\,\times \exp\left(-i\int_0^t J_\mu(X(\sigma))dX^\mu(\sigma)\right)\bigg\}\tag{5}
\end{align*} 

Unfortunatelly eq(4)-eq(5) still are somewhat mathematically formal and evaluations with them has never been performed in the literature, even if on the non-relativistic case it leads to the correct results. However it is worth to remaind that on lattice, the chances for a fully mathematical rigorous and calculational scheme are greater than its version on the continuum $\mathbb{R}^D$. This vital matter will be presented elsewhere.

At this point, it is instructive to point out the supersymmetric path integral proposal for a spinning particles ([4],[5]). However its suggested Wilson Loop necessarilly would involves the spin orbit coupling term with the strenght field which is loop lenght dependent. Explicitly:
\begin{align*}
&W^B[C_{xx}] = {\rm Tr}_{SU(N)} \bigg\{{\rm Tr}_{\rm Dirac} \bigg(\mathbb{P}_{\rm Dirac} \bigg(\mathbb{P}_{SU(N)}\\
&\,\,\,\exp\bigg[-i\int_0^t \big(A_\mu(X(\sigma))\dot{X}^\mu(\sigma) + \frac 14\, i[\gamma^\mu,\gamma^\nu]F_{\mu\nu}(X(\sigma)d\sigma\big)
\bigg]\bigg)\bigg\} \tag{6}
\end{align*}

However to reformulate the ill defined quantum field theory of $Q.C.D(SU(N))$, one makes the hypothesis that the Wilson Loop on eq(5) is the correct collective variable to be represented through a random surface path integral at least on the lattice framework ([1]). We point out that this step can be regard as a correct ``theoretical physics" argument at the deep infrared region, where the Dirac spin degrees of freedom are  frozens. Note that on light of this hypothesis, the string path integral representing the quantum Wilson Loop on eq(5) must be supersymetrized on the string ambient space-time to fully represent the spinning Wilson Loop eq(6) ([4],[5]). But mathematical proofs are required and have not been available since still there is not a mathematical theory for spinning Brownian motion at the present time. ([6])

But the whole point of finding string representations for Q.C.D$(SU(N))$ (or\linebreak $Q.C.D(SU(\infty))$)more precisely  is to argue that Q.C.D as a mathematical object is an ill-defined object by itself. So all previous formulae eq(1)-eq(6) are only suggestive at the continuum and could mathematically make sense (if any) only at lattice, where $Q.C.D(SU(\infty))$ is well-defined. As a result one must search a string path integral from a formal point of view, heavily inspired on the formal objects eq(1)-eq(6). And Q.C.D should be fully replaced by the string path integral, which must reproduces lattice Q.C.D, \textit{when reformulated in the lattice${}^*$} \footnote{${}^*$ lattice = computer} (\textit{the string path integral}!). It is even expected that that Q.C.D string path integral on lattice is the calculational tool for Q.C.D evaluations. A final remark: $Q.C.D(SU(N_c))$ string path integral is expected (but not proved yet!) to be the possible found $Q.C.D(SU(\infty))$ now endowed with all non trivial genus -- Unitarization of the associated Q.C.D's Scattering Matrix). However, the determination of $N_c=3$ must be make recourse to the flavor quark electro-weak sector and to the Baryons excitations. All of this surely ``sconisciuta terra".

So let us use scalar deep infrared $Q.C.D(SU(\infty))$ (where the Yang-Mills quantum average are granted to be factorized on the product of gauge invariant observables).

In this case, we have written the following loop wave string like wave equations  eq(7-b) for formal Euclidean Yang-Mills theory under the hypothesis of a non-zero Yang-Mills strenght condensate ([7],[8]). Namely (see Appendix 1)
\begin{equation}
\Phi_{N_c}[X_\mu(\sigma),\,0\le\sigma\le2\pi] = \frac{1}{N_c} \bigg[{\rm Tr}_{SU(N_c)} \bigg\{ \mathbb{P}\left(\exp i \int_0^{2\pi} A_\mu(X(\sigma))dX^\mu(\sigma)\right)\bigg\}\bigg] \tag{7-a}
\end{equation}

\begin{align*}
\int_0^{2\pi} &d\bar\sigma \left[\left(\frac{\delta^2}{\delta X_\mu(\bar\sigma)\delta X^\mu(\bar\sigma)}\right) - \big\langle{\rm Tr}_{SU(\infty)} (F^2)\big\rangle |\dot{X}_\mu(\bar\sigma)|^2\right]\Phi_\infty[X_\mu(\bar\sigma)]\\
&= (g^\infty)^2 \bigg\{\int_0^{2\pi}d\sigma \int_0^{2\pi} d\bar\sigma\, \delta^{(D)}(X_\mu(\bar\sigma)-X_\mu(\sigma))\times (\dot{X}_\mu(\sigma) \dot{X}_\mu(\bar\sigma))\\
&\qquad\qquad\quad\Phi_\infty[X_\mu(\tilde\sigma); 0 \le \tilde\sigma \le \bar\sigma] \Phi_\infty[X_\mu(\tilde\sigma); \bar\sigma\le\tilde\sigma\le 2\pi] \bigg\}\tag{7-b}
\end{align*}

\vglue .4in

\noindent{\large\bf 1.3\, The ``free" string path integral}

\vglue .2in

In order to search solutions for the non linear ``quadratic" loop space wave equation eq(7-b), one must give a correct meaning for the Wiener-Feynman sum under surfaces (bosonic Brownian surfaces), in place of the well known bosonic sum over Wiener-Feynman paths.

One fashionable proposal is due to A.M. Polyakov ([5]), altought it has been revealed to be clearly wrong (Appendix 2).

It is based on the Brink-Howe action, but added with a non-vanishing cosmological term
\begin{align*}
G(C) &= \int d_\mu^{\rm cov}[g_{ab}(\xi) \int_{\partial X^\mu\underbrace{(\sigma,t)}_{\xi} = C^\mu(\sigma)} d^{\rm cov}[X_\mu(\xi)]\\
&\exp\bigg\{-\frac{1}{\pi\alpha'} \int_{\mathbb D} \left(\frac 12\, \sqrt g\, g^{ab}\,\partial_aX^\mu\,\partial_b X_\mu\right)(\xi)d^2\xi\\
&\qquad\quad -\mu_0^2 \int_{\mathbb D} (\sqrt g)(\xi)d^2\xi\bigg\} \tag{8}
\end{align*}

However in order for the classical solutions of eq(8) reproduces the Nambu-Goto area functional it is going to constraint the cosmological term to vanish (unless at the extrinsic space-time dimension $D=2$, a two restrictive dimensionality for the space time quantum dynamics).

The full correct meaning of eq(8) was however written in full in ref([9])
\begin{align*}
G(C_{xx}) &= \int d^{\rm cov}_\mu[g_{ab}(\xi)] \int_{\partial X^\mu(\sigma,\zeta)=C_{xx}^\mu(\sigma)}
d^{\rm cov}[X^\mu(\xi)]\\
&\exp\left\{-\frac 12 \int_{\mathbb D} (\sqrt g\,g^{ab}\,\partial_a X^\mu\,\partial_b X_\mu)\right\}\\
&\exp\left\{-\left(\frac{1}{2\pi\alpha'}\right) \int_{\mathbb D} (\sqrt g)(\xi)d^2\xi\right\}\\
&\delta_{\rm cov}^{(F)} \left(g_{ab}(\xi) - \big(\partial_a X^\mu\, \partial_b X_\mu\big)(\xi)\right) \tag{9}
\end{align*}

It results that on the surface conformal gauge and for $D=26$ (or by introducing $N$ neutral fermions such that $N+D=26$), one obtains the expected gauge fixed propagator as a theory of free fields on the string domain parameter (otherwise the theory is non-renormalible, so ill-defined -- see also Appendix 2).
\begin{align*}
G(C_{xx}] &= \int_{\partial X^\mu(\xi)=C_{xx}^\mu} D^F(X^\mu(\xi)]\times \delta^{(F)}\big((\partial_+X^\mu)^2+(\partial_-X^\mu)^2\big)\\
&\quad\exp \left[-\frac{1}{2\pi\alpha'} \int_{\mathbb D} \big((\partial_+X^\mu)(\partial_-X_\mu)\big)(\xi^+,\xi^-)d^2\xi\right] \tag{10}
\end{align*}

At this point it is argued ([1],[8],[10]) that the following two-dimensional path integral, with a neutral set of $N=22$ fermions solve the $(Q.C.D(SU(\infty))$ loop wave equation\linebreak eq(7-b) (with $\langle 0|F^2|0\rangle_{SU(\infty)} = \dfrac{1}{\pi\alpha'}=1$ and $\xi = (\sigma,\tau)$).
\begin{align*}
&\Phi_{SU(\infty)} \left[C_\mu(\sigma),\, 0 \le \sigma \le t\right] = \int_0^\infty dA\bigg\{\int_{\partial X^\mu(\xi)=C^\mu} D^F[X^\mu(\xi)]\\
&\exp\left[-\frac 12 \int_0^A d\tau \int_0^t d\sigma(\partial X^\mu)^2(\sigma,\tau)\right] \int(D\psi^i\,D\overline{\psi}^i)(\xi)\\
&\exp\left[+\int_0^A d\tau \int_0^t d\sigma \big(\sqrt{h(X)}\,\overline{\psi}^i \partial_h\,\psi^i\big)(\sigma,\tau)\right]\\
&\exp\bigg[-\lambda_0^2 \int_0^A d\tau \int_0^A d\tau' \int_0^t d\sigma \int_0^t d\sigma' \big(\sqrt{h(X)}\,\overline{\psi}^i\,\psi_i\big)(\sigma,\tau)\\
&\big(\sqrt{h(X)}\,\overline{\psi}^i\,\psi_i\big)(\sigma',\tau') \mathfrak{I}^{\mu\nu}(X(\xi)\delta^{(D)}(X(\xi)-X(\xi'))\mathfrak{I}_{\mu\nu}(X(\xi'))\bigg]\bigg\} \tag{11}
\end{align*}

Here $\mathfrak{I}^{\mu\nu}(X(\xi))$ is the normalized surface area tensor (see footnote). 

\footnote{Firstly one must imposes the ``Pauli-Fermi" conditions on the possible Q.C.D string surface $\mathfrak{I}^{\mu\nu}(X(\sigma,\tau)) \mathfrak{I}_{\mu\nu}(X(\sigma',\tau))=0$ for $\sigma \ne \sigma'$. For trivial self intersection points $(\sigma,\tau) = (\sigma',\tau')$ one has formally the result $\delta^{(D)}\big(X_\mu(\xi)-X_\mu(\xi')\big) = \dfrac{\delta^{(2)}(\xi-\xi')\delta_\varepsilon^{(D-2)}(0)}{2^{D/2}\cdot h^{D/8}(X(\xi))}$. So one can expect that for $D=4$, the self avoiding term (responsible for the $Q.C.D(SU(\infty))$ string be an full interacting string theory) reduces to a $U(11)$ Gross-Neveu $2D$ model on the string surface.}

\begin{equation}
\mathfrak{I}^{\mu\nu}(X(\xi)) = \frac{(\varepsilon^{ab}\partial_aX^\mu \partial_b X^\nu)(\xi)}{\sqrt 2\,\sqrt{h(X(\xi))}} \tag{12-a}
\end{equation}

\medskip

\begin{equation}
\mathfrak{I}^{\mu\nu}(X(\xi)) \mathfrak{I}_{\mu\nu}(X(\xi)) = 1 \tag{12-b}
\end{equation}

\medskip

At this point one could consider eq(11) as an interacting string path integral on the surface conformal gauge as done in eq(10).

Note that at this point it is somewhat irrelevant to consider the two-dimensional path integral as a path integral related to a random surface theory, even if this geometrical interpretation holds true on lattice, and necessary for theory's unitarization afterwards 

Q.C.D is thus analitically solved through interpreting eq(11) as a string path integral extended to all surface genus (somewhat related the Mandelstam  light-cone string path integral on euclidean space-time).

At this point of ours comments, we remark that eq(11) should be evaluated explicitly in terms of the loop boundary $C_{xx}$ and the loop parameter $\sigma$ and the string proper-time $A$ (see [8] for details on a simple model of constant gauge fields configurations). After this step, one expects that this ``stringy" Wilson Loop is now well-defined and should replace the ill defined one given by eq(7-a) and averaged over the (ill defined) quantum euclidean Yang-Mills path integral. Namely
\begin{align*}
&\left\langle\frac{\det \not{\!\!D}(A_\mu+J_\mu)}{\det \not{\!\!D}(0)}\right\rangle_{SU(\infty)} =
\exp\bigg\{-\int_0^\infty \frac{dt}{t} \int d^\nu x \bigg\{\int_{C_{xx}^\mu(\sigma)]} D^F[C_{xx}^\mu(\sigma)]\\
&\int_{\pi^\mu(0)=\pi^\mu(t)} D^F[\pi^\mu(\sigma)]\exp\left(i \int_0^t \pi^\mu(\sigma)\dot{X}_\mu(\sigma)d\sigma\right)\\
&{\rm Tr}_{\rm Dirac} \left[\mathbb{P}_{\rm Dirac}\bigg\{\exp\left(-i\int_0^t (\gamma^\mu \pi_\mu)(\sigma)d\sigma)\right)\right]\\
&\exp\left(-i \int_0^t J_\mu(X(\sigma))dX^\mu(\sigma)\right) \times \Phi_{SU(\infty)} [C_{xx}^\mu(\sigma),\,0 \le \sigma\le t]\bigg\} \tag{13}
\end{align*}

We have thus that eq(13) should be the correct (string) definition of $Q.C.D(SU(\infty))$ or possible $Q.C.D(SU(N_c))$ when the surface sum is defined for all possible topological genera and adjusted to the eletroweak sector of Nuclear Forces ([11]).

It is thus suggested by eq(13) that correlation functions of the quarks color singlet bilinear in the ill defined Lagrangean quantum field Q.C.D are well defined by the (on shell) string vertexs averages scattering amplitudes associated to eq(11), producing as a result, the meson $S$-matrix and the determination of it physical spectrum (the meson mass spectrum through the Regge dual model predictions. The most relevant basic problem on extending successfully quantum mechanics for Elementary Particle Physics).

So one must use the possible well-defined Q.C.D string theory in place of the quantum field ill defined Q.C.D Lagrangean. Both coinciding only at lattice as well-defined mathematical Quantum Euclidean theories.

In this context of $Q.C.D(SU(\infty))$ meson spectrum, one should point out that the numerical condition of vanishing of the conformal anomaly in the free string propagator eq(9) or in the proposed $Q.C.D(SU(\infty))$ eq(11), $26=D$, or $26=D+N$ respectivelly is replaced by taking the string two-dimension parameter Planck's constant 
$(\hbar){^(2)}$ to vanishes, which means that $(\partial_+X^\mu\,\partial_- X_\mu)(\xi) \sim h_{ab}^{\rm classical}(X^\mu(\xi)) =\dfrac {(d\sigma^2)+(d\tau)^2}{\tau^2}$\, ([12]).

Work on the space-time supersymmetric version of eq(11) will appear  elsewhere. (See Appendix 3).

\medskip

\noindent{\large\bf Acknowledgments:} We are thankfull to CNPq for a Senior Pos Doctoral Fellowship.

\vglue .2in

\noindent{\large\bf References}

\medskip

\begin{itemize}
\item[{[1]}] Luiz C.L. Botelho - Journal of Mathematical Physics, vol. 30, 2160, (1989).
\item[{-}] Rev. Bras. Fis., vol. 16, p. 279, (1986).
\item[{-}] Caltech Preprint (1987).
\item[{[2]}] A.A. Migdal - Nucl Phys B, vol. 189, p. 253, (1981).
\item[{[3]}] A.M. Polyakov - Nucl Phys, vol. 
B486, p.23, (1997).
\item[{-}] Luiz C.L. Botelho - Modern Phys. Letters B, vol. 13. n. 687, p. 203, (1999).
\item[{-}] A.I. Karanikas and C.N. Ktorides - Phys Lett 235B, vol. 235, p.90, (1990).
\item[{[4]}] Luiz C.L. Botelho - Phys. Letters 169B, 428, (1986).
\item[{-}] S.G. Rajeev - Annals of Physics 173, p.249, (1987).
\item[{[5]}] A.M. Polyakov - ``Gauge Field and Strings", Harwood Academic Chor, Switzerland, (1987).
\item[{[6]}] Luiz C.L. Botelho - Random Operators and Stochastic Equations, vol. 21, p. 271, (2013).
\item[{[7]}] Luiz C.L. Botelho - International Journal of Modern Physics A, vol. 32, 1750030, (2017).
\item[{[8]}] Luiz C.L. Botelho - International Journal of Modern Physics A, vol. 32, 1750031, (2017).
\item[{[9]}] Luiz C.L. Botelho - Phys. Rev. 49D, 1975, (1994).
\item[{[10]}] Luiz C.L. Botelho - International Journal of Theoretical Physics, vol. 48, 2715, (2009).
\item[{[11]}] Luiz C.L. Botelho - ``Methods of Bosonic and Fermionic Path Integrals Representations - Continuous Random Geometry in Quantum Field Theory, Nova Science Publishers, ISBN 578-1-60456-068-8, (2009)
\item[{[12]}] Luiz C.L. Botelho - Lecture Notes in Topics in Path Integrals and String Representations, World Scientific Publishing, ISBN 9889813143463, (2017).
\item[{[13]}] Lars V. Ahlfors - Complex Analysis, Third Edition - McGraw-Hill International Editions, 1979.
\item[{[14]}] S.G. Mikhlin - Integral Equations, Pergamon Press - Pure and Applied Mathematics, 1957.
\end{itemize}

\newpage

\centerline{\large\bf Appendix 1}

\medskip

\centerline{\large\bf The $\pmb{Q.C.D(SU(\infty))}$ Loop Wave Equation}

\vglue .3in

It has been fully discussed in the literature that after formal manipulations on the objects involved, specially with the (ill defined) Yang-Mills path integral in the continuum, one arrives at the following (formal) loop wave equation for the Wilson (see eq(7-a) in the text) loop in 
$Q.C.D(SU(N))$:
\begin{align*}
\int_0^{2\pi}
 &d\bar\sigma \bigg\{\frac{\delta^2}{\delta X_\mu(\bar\sigma)\delta X^\mu(\bar\sigma)} - \left(\int d^D x \langle\Omega|F_{\mu\nu}(x) F_{\mu\alpha}(x)|\Omega\rangle_{SU(N_c)}\right) \frac{d X^\nu(\bar\sigma)}{d\bar\sigma}\,\frac{d X^\alpha(\bar\sigma)}{d\bar\sigma}\bigg\}=\\
&(g^2N_c) \bigg\{\int_0^{2\pi} d\sigma \int_0^{2\pi} d\bar\sigma\,\delta^{(D)}(X^\mu(\sigma)-X^\mu(\bar\sigma))\left(\left\vert\frac{dX^\beta(\sigma)}{d\sigma}\,\frac{dX_\beta(\bar\sigma)}{d\bar\sigma}\right\vert\right)\\
&\left\langle\Phi_{N_c}[X_\mu(\tilde\sigma); \,0 \le \tilde\sigma \le \bar\sigma] \times 
\Phi_{N_c}[X_\mu(\tilde\sigma);\, \bar\sigma \le \tilde\sigma \le 2\pi]\right\rangle_{SU(N_c)}\bigg\} 
\tag{1-1}
\end{align*}
where $\langle\quad\rangle_{SU(N_c)}$
denotes the average over the (ill defined) Yang-Mills euclidean path integral.

As it has been observed in the main text, the correct mathematical meaning of eq(1-1) should be searched on the well-define lattice formulation of Q.C.D.

However one defines a ``zeroth-order" for the full stochastic-Dhyson Schwinger equation through the crude force approximation, called by us of $Q.C.D(SU(\infty))$, where the Yang-Mills path integral of the gauge invariant observables factorize, as a mathematical methods working calculational hypothesis. 

It has argued in the literature that such approximation on the continuum formal Yang-Mills theory should be justified by the well-known t'Hooft diagrammatic framework ([2]). However this suggestion has not been fully proved in our opinion. The loop equation eq(1-2) should be better regarded perhaps in the framework of Random Matrix Theory.

As a result one gets eq(7-b) written in the main text.

Note that we have made the simplifying assumption of the space-time isotropy of the non perturbative Yang-Mills strenght field condensate on eq(1-1) (see refs([2],[8]). Otherwise one must consider the $Q.C.D(SU(\infty))$ free string moving in a sort of metric back ground, namely ([10]):
$$
G^{(0)}(C_{xx}) = \int_{\partial X^\mu=C_{xx}} D^F[X^\mu(\xi)]\exp\left\{-\frac 12 \int_\xi \big[(\partial_t X^\mu)^2+G_{\mu\nu}\partial_\sigma X^\mu \partial_\sigma X^\nu\big](\sigma,\zeta)\right\}
$$
with
$$
G_{\nu\alpha} = \int d^Dx \big\langle\Omega |(F_{\mu\nu}\,F_{\mu\alpha})(x)|\Omega\big\rangle_{SU(\infty)}
$$

\vglue .9in

\centerline{\large\bf Appendix 2}

\vglue .2in

\centerline{\large\bf The Correct Free String Propagator In Loop Space 
$\pmb{Q.C.D(SU(\infty))}$}

\vglue .3in

Let us sketchy the anomaly evaluation of the string path integral eq(9) in the bulk of this note.

\vglue .2in

The main point is by following closely the detailed proof as given by us in our previous work ([9]) to fix the dipheomorfism gauge to the conformal gauge both on the path integral over the string vector position and its Riemaniann structure also (both regarded a priori independent), but related by the covariant delta constraint integral in the path integral measure. Namely ([7],[8])
\begin{equation}
X^\mu(\xi) = X^{\mu,CL}(\xi) + \sqrt{\pi\alpha'}\,Y^\mu(\xi) \tag{2-1}
\end{equation}
\begin{equation}
\left(\Delta_{h(X^{\mu,CL)}}\right) X^{\mu,CL} = 0 \tag{2-2}
\end{equation}
\begin{equation}
\partial X^{\mu,CL} = C_{xx}^\mu \tag{2-3}
\end{equation}
\begin{equation}
g_{ab}(\xi) = h_{ab}\big(X^{\mu,CL}(\xi)\big) \beta^2(\xi) \equiv h_{ab}^{CL}(\xi)\beta^2(\xi) \tag{2-4}
\end{equation}

\vglue .2in

As a result one obtain the Liouville model in the sigma model like form with the usual Feynmann product measure:

\newpage

\begin{align*}
G(C_{xx}) &= \int \prod_{(\xi)} \big(\sqrt{h^{CL}(\xi)}\,d\beta(\xi)\big)\\
&\times \bigg\{ \prod_{(\xi)} 
\big(\sqrt[\uproot{2} 4]{h^{CL}(\xi)}\,dY^\mu(\xi)\big)\\
&\exp\bigg\{+\frac{(26-D)}{4\delta\pi} \int d^2\xi\left(\sqrt{h_{CL}}\, h_{CL}^{ab}\, \partial_a(\beta)\partial_a\left(\frac{1}{\beta}\right)\right)(\xi)\bigg\}\\
&\times \exp\bigg\{-\frac{\mu_R^2}{2} \int d^2x \big(\sqrt{h_{CL}}\,.\,\beta^2\big)(\xi)\\
&\times \exp\bigg\{-\frac{1}{2\pi\alpha'} \int d^2\xi \big(\sqrt{h_{CL}}\,h^{+-}_{CL}\,\partial_+Y^\mu\,\partial_- Y_\mu\big)(\xi)\bigg\}\\
&\times \exp\bigg\{\int d^2\xi \big(\sqrt{h_{CL}}\,R(h_{ab}^{CL})\,.\,\beta\big)(\xi)\bigg\} \\
&\times\delta^{(F)}_{{\rm cov}}\big[\big(\beta^2(\xi) h_{+-}^{CL} - \partial_+ Y^\mu \partial_-Y_\mu\big)(\xi)\big]
\tag{2-5}
\end{align*}

Since the above two-dimensional model is non-renormalizable after the realization of the  path integral over the $\beta(\xi)$ field (due to the delta functional constraint eq(2-5), one must vanishe the conformall anormaly factor term by choosing $D=26$. At this point it is worth to remark that $\mu_R^2 = \lim\limits_{\varepsilon\to 0} 
\dfrac{(2-D)}{\varepsilon}$ and should be considered as just a formal renormalization of the Regge constant $\mu^2 = \dfrac{1}{\pi\alpha'}$\,.

It is worth also to note that non trivial topology is straightforwardly taken into account through the classical minimal surface equation eq(2-2)-eq(2-3), now a Dirichlet problem (only at the surface conformal gauge ([7],[8]) to be solved at a Riemann surface of arbitrary genus (see ref[13] -- Chapter 6 -- Section 5: canonical conformal mappings of multiply connected regions; and ref[14] -- Chapter IV - \S 31 -- Dirichlet's problem for multi-connected regions).

It may be also solved by the introduction of the Surface's Riemann Structure Techmilles parameter directly on the decomposition eq(2-4) ([11]).

The resulting anomaly free and in the euclidean light cone gauge fixed string path integral is the well-known Mandelstan string path integral in the ``slites" string parameter domain $\xi$.

\newpage

\vglue .8in

\centerline{\large\bf Appendix 3}

\vglue .2in

\centerline{\large\bf The space-time supersymmetric self avoiding string}

\vglue .3in

Let us call attention that in this case the fermions random surface oriented tensor has as  candidate the following expression (non normalized to unity)
\begin{equation}
\mathfrak{I}_{\mu\nu}^F \big(X^{\alpha,F} (\xi,\theta)\big) \equiv \left(\frac{\varepsilon^{ab}\partial_aX_\mu\,\partial_bX_\nu + \psi^\mu(\gamma_a\partial_b)\psi^\nu}{\big[\det [H_{ab}]\big]^{1/2}}\right)(\xi) \tag{3-1}
\end{equation}
where the fermionic surface Riemann matrix is explicitly given by $(\xi = (\sigma,\tau))$
\begin{equation}
H_{ab}(\xi) = \big(\partial_aX^\mu\,\partial_bX_\mu + \psi^\mu(\gamma_a\partial_b)\psi_\mu\big)(\xi) \tag{3-2}
\end{equation}

The Nambu-Goto area functional is thus given by
\begin{equation}
S^{(0)} = \frac{1}{2\pi\alpha'} \int d^2\xi \big[\det[H_{ab}]\big]^{1/2}(\xi) \tag{3-3}
\end{equation}

The self suppressing term is expected to be given by
\begin{align*}
S^{(1)} &= \int d^2\xi \int d^2\xi' \int d\theta \int d\theta'[(\overline{\psi}\psi)(\xi)\,(\overline\psi\psi)(\xi')]\\
&\mathfrak{I}_{\mu\nu}^F(X^{\alpha,F}(\xi,\theta)) \delta^{(D)}(X_\mu^F(\xi,\theta)-X_\mu^F(\xi',\theta') \times \mathfrak{I}_{\mu\nu}^F (X^{\alpha,F}(\xi',\theta')) \tag{3-4}
\end{align*}

\newpage

\centerline{\large\bf Appendix 4}

\vglue .3in

In this appendix we intend to write the classical string representation for the\linebreak $Q.C.D(SU(\infty))$ euclidean quantum Wilson Loop a suggested by eq(7-a) for $26=D+N$ and frozen internal fermionic degrees of freedom.

By solely considering the classical light-cone gauge fixed minimal surface $X^{\mu,CL}(\xi)$ and defined for a domain string parameter of arbitrary topological genera (see remarks below eq(2-5) -- Appendix 2). One has thus the representation below:
\begin{align*}
\Phi_\infty^{\rm Semi-Classical} [C_{xx}]\, &\sim
\int_{M^{\text Teic}(h)} d\nu(\tau_i) \exp \bigg\{-\frac{1}{2\pi\alpha'} \int_D d^2\xi\big(\sqrt h\,h^{+-}\big)(\tau_i,\xi)\\
&\times \partial_+\,X^{\mu,CL}(\xi,\tau_i)\partial_-\,X_\mu^{CL}(\xi,\tau_i)\bigg\} \tag{4-1}
\end{align*}

Here $h_{ab}(\xi,\tau_i)$ are the metric representatives on the Teichmüller space of the possible topological structure on the metric space of the surface $2D$ manifold $\{X^{\mu,CL}(\xi)\}$ and $d\nu(\tau_i)$ denotes the measure the integration of such Teichmüller space of dimensionality $6g-6$ where $g$ is the string surface genus. The exact expression of such measure $d\nu(\tau_i)$ is explicitly given on ref [12] (reference number 15 on this pointed out ref [12]).

However a more suitable Q.C.D string effective representation path integral can be given by U(11) Gross-Neveu model defined over a classical Riemannian surface of genus $g$ and summed up over all theses classical Riemann surfaces with the associated 
Nambu-Goto height area functional, but with the path integral defined only over the surface's Teichmüller parameters.

Note that this classical Riemann surface is explicitly given by eq(2-1)-eq(2-4) of\linebreak Appendix 2. This result thus lends perhaps to the loop space-string integrability of $Q.C.D(SU(\infty))$

\end{document}